\documentclass[10pt]{iopart}
\usepackage{cite}

\begin{document}

\title{Star-product quantization and symplectic tomography}

\author{Olga V. Manko}

\address{P.\,N.\,Lebedev Physical Institute, Leninskii prospect 53, Moscow 119991, Russia}
\ead{omanko@sci.lebedev.ru}
\begin{abstract}
A review of the symplectic tomographic approaches within the
framework of star-product quantization is presented. The
classical statistical mechanics within the framework of the
tomographic representation is considered. The kernels of star-product of functions
- symbols of operators in classical and quantum mechanics are
presented.\\
PACS number(s): 42.50.-p, 03.65.Bz

\end{abstract}

\section{Introduction}
In quantum mechanics, the state is described by the wave function or
density matrix, the observables are described by operators acting in
Hilbert space.

In classical mechanics, the state of a system with fluctuations is
described by the probability distribution function and
observables are described by functions.

So, we have different languages for describing the quantum and classical
nature, but to understand the nature of the system consisted of the
classical and quantum parts, it is necessary to have the same language
for both domains. Due to this, an idea appeared to create the probability
representation for quantum mechanics along with the probability representation
for the classical states.

The tomographic probability representation of quantum mechanics was
introduced in \cite{Mancini1,Mancini2,Mancini3} and the tomographic probability
representation of the classical states was introduced in
\cite{OMVI,MankoManko1,MankoManko}. In the probability representation,
the quantum and classical states are described by the same
objects -- tomograms. Tomograms are positive measurable
probability-distribution functions of random variables, which are
determined in an ensemble of the reference frames in the system's phase
space.

In order to describe observables by functions instead of
operators, in quantum mechanics the quantization based on
star-product of functions is used. In \cite{Marmo,Kapusta} it was
shown that symplectic tomography scheme (probability representation)
is a new example of quantization based on star-product of functions
-- symbols of operators. The operators determining the star-product
quantization scheme and the kernel of star-product of symbols of
operators for symplectic tomography were obtained in the explicit form.
The tomographic symbols of classical observables were discussed in
\cite{MankoManko}. The tomographic star-product kernel in classical
mechanics was obtained in \cite{Pilyavets}. A scheme of quantization
dual to symplectic tomography scheme and its connection  with different
nature of the density operators and operators--observables were discussed
in \cite{Patricia}.

The aim of this paper is to present a review of the symplectic
tomographic approaches in both the quantum and classical domains
within the framework of star-product quantization.

\section{Classical states}
Let us consider a particle with one degree of freedom with unit
mass. The position and velocity of the particle are $-\infty<q<\infty$
and $-\infty<\dot q<\infty$, respectively. The particle momentum
is $p=\dot q$ due to $m=1$. The particle state is identified with a point
in the phase space (plane) with coordinates $q$ and $p$. The evolution of
the particle state is described by a trajectory in the phase space
$q(t),p(t)$.

Let us suppose that the classical particle is located inside
some environment and the position $q$ and momentum $p$ fluctuate. In
view of these fluctuations, the particle state is described by a
probability distribution function $f(q,p)$, which is nonnegative
$f(q,p)\geq 0$ and normalized
\begin{equation}\label{eq.2}
\int f(q,p)\,d q\,d p=1.
\end{equation}
We consider a point in the phase space $(p,q)$. If one rotates the
reference frame in the phase space by an angle $\phi$ and then makes
the scaling transformation, the coordinate of the point in the new
reference frame is
\begin{equation}\label{eq.1} X\left (q,\,p\right )=\mu
q+\nu p,\end{equation} where $\mu=s\cos\phi$ and $\nu=s^{-1}\sin\phi.$
The parameters $\mu$ and $\nu$ determine the reference frame.

The tomogram of this state is determined by the Radon transform of the
probability distribution function $f(q,p)$ as follows:
\begin{equation}\label{eq.1a}
w_f\left (X,\,\mu ,\,\nu \right )=\int f_{\mbox{cl}}\left
(q,\,p\right )\, \delta \left (\mu q +\nu p -X\right )\,dq\,dp,
\end{equation}
and it is called classical tomogram. Classical tomogram can be
determined as the expectation value of a delta-function calculated
with the help of the distribution function $f(q,p)$ in the phase space
\begin{equation}\label{eq.1b} w_f\left (X,\,\mu ,\,\nu \right )=\langle \delta
\left (\mu q +\nu p -X\right )\rangle.
\end{equation}
The classical tomogram is nonnegative
$w_f\left (X,\,\mu ,\,\nu \right )\geq0$,
normalized
$\int w_f\left (X,\,\mu ,\,\nu \right )\,d X=1$, and a homogeneous function
$w_f\left (\lambda X,\,\lambda\mu ,\,\lambda\nu \right
)=|\lambda|^{-1}w_f\left (X,\,\mu ,\,\nu \right ).$

The physical meaning of the classical tomogram is that it is the
probability density for the particle coordinate $X$, which is
measured in the phase-space reference frame subjected to scaling
of the axes and subsequent rotation with respect to the original
reference frame. In the same way, as in the case of Fourier
transform, where information contained in the function
is equivalent to information contained in its Fourier transform,
information on the particle state contained in the distribution
function $f(q,p)$ is equivalent to information contained in its
Radon transform -- the classical tomogram $w_f\left (X,\,\mu ,\,\nu
\right )$. The Radon transform is invertible
\begin{equation}\label{eq.1c}
\fl f_{\mbox{cl}}\left (q,\,p\right)=\frac {1}{4\,\pi^2}\int
w_f\left (X,\,\mu ,\,\nu \right)\exp \left [-i\left (\mu q+\nu p
-X\right ) \right ]\,dX\,d\mu \,d\nu;
\end{equation}
thus, knowing the classical tomogram $w_f\left (X,\,\mu ,\,\nu \right)$,
one can reconstruct the probability distribution function $f(q,p)$.

\section{Quantum states}
In quantum mechanics, the Wigner function \cite{Wigner32} plays the role
of the probability distribution function. It was
shown~\cite{Mancini1, Mancini2} that for the  generic linear
combination of quadratures, which is a measurable observable $\left
(\hbar =1\right)$
\begin{equation}\label{X}
\widehat X=\mu \hat q+\nu\hat p,
\end{equation}
where $\hat q$ and $\hat p$ are the position and momentum,
respectively, the function $w\,(X,\,\mu,\,\nu )$ (normalized with
respect to the variable $X$), depending on the two extra real
parameters $\mu $ and $\nu ,$ is related to the state of the quantum
system expressed in terms of its Wigner function $W(q,\,p)$ as
follows:
\begin{equation}\label{w}
\fl w\left (X,\,\mu,\,\nu \right )=\int \exp \left [-ik(X-\mu q-\nu
p)\right ]W(q,\,p)\,\frac {dk\,dq\,dp}{(2\pi)^2}\,.
\end{equation}
This function was called symplectic tomogram.  The physical meaning
of the parameters $\mu $ and $\nu $ is that they describe an
ensemble of rotated and scaled reference frames in which the
position $X$ is measured.

For $\mu =\cos \,\varphi $ and $\nu =\sin\,\varphi ,$ the marginal
distribution~(\ref{w}) is the distribution for the homodyne-output
variable used in optical tomography~\cite{Ber,VogRi}.

Formula~(\ref{w}) can be inverted and
the Wigner function of the state can be expressed in terms of
symplectic tomogram~\cite{Mancini1}
\begin{equation}\label{W}
\fl W(q,\,p)=\frac {1}{2\pi }\int w\left (X,\,\mu ,\,\nu \right )
\exp \left [-i\left (\mu q+\nu p-X\right )\right ] \,d\mu \,d\nu
\,dX\,.
\end{equation}
Since the Wigner function determines completely the quantum state of
a system and, on the other hand, this function itself is completely
determined by symplectic tomogram, one can use for describing
quantum states symplectic tomograms (positive and normalized) which
are probability distribution functions analogous to the classical ones.

The quantum state is given if the position probability distribution
$w\left(X,\,\mu,\,\nu \right)$ in an ensemble of rotated and
squeezed reference frames in the phase space is given. Information
contained in symplectic tomogram $w\left(X,\,\mu,\,\nu \right)$ is
overcomplete. To determine the quantum state, it is enough to know
the values of symplectic tomogram of the state for arguments
satisfying the condition
$\left(\mu^2+\nu^2=1\right)$,
where $\mu=\cos \varphi$. This representation of quantum mechanics
called probability representation was introduced in \cite{Mancini1,
Mancini2,Mancini3}. In \cite{Ariano} it was shown that the density
operator can be reconstructed from symplectic tomogram
\begin{eqnarray*}
\hat\rho=\frac{1}{2\pi}\int w(X,\mu,\nu)e^{i(X-\mu\hat q-\nu\hat
p)}\,d X\, d\mu \,d\nu. \end{eqnarray*}

\section{General star-product scheme}
In quantum mechanics, the observables are described by operators
acting in Hilbert space of the states. In order to consider the
observables as functions in a phase space, first we describe following
\cite{Marmo,Kapusta} a general construction of a map of the
operators onto the functions without any concrete realization of the
map.

We consider an operator $\hat A $ acting in a given Hilbert
space. Let us suppose that we have a set of operators $\hat{\cal
U}({\bf x})$ acting in the Hilbert space $H$, where the
$n$-dimensional vector ${\bf x}=(x_1,x_2,\ldots ,x_n)$ labels the
particular operator in the set. We construct the c-number function
$f_A({\bf x})$ using the definition
\begin{equation}\label{eq.1}
f_{\hat A}({\bf x})=\mbox{Tr}\,(\hat A\hat{\cal U}({\bf x})).
\end{equation}
The function $f_A({\bf x})$ is called the symbol of operator $\hat
A$, and the operators $\hat{\cal U}({\bf x})$ are called dequantizers
\cite{Patricia}. We suppose that this relation has an inverse.
There exists the set of operators $\hat{\cal D}({\bf x})$ acting in
the Hilbert space such that
\begin{equation}\label{eq.2}
\hat A= \int f_{\hat A}({\bf x})\hat{\cal D}({\bf x})~d{\bf x}.
\end{equation}
The operators $\hat{\cal D}({\bf x})$ are called quantizers
\cite{Patricia}. The formulae are selfconsistent if the
following property of the quantizers and dequantizers takes place:
\begin{equation}\label{SVs15}
\mbox{Tr}\left[\hat{\cal U}({\bf x}) \hat{\cal D}({\bf x}')\right]=
\delta\left({\bf x}-{\bf x}'\right).
\end{equation}
Relations~(\ref{eq.1}) and~(\ref{eq.2}) determine the invertable
map of the operator $\hat A$ onto the function $f_{\hat A}({\bf x})$.

The most important property is the existence of associative product
(star-product) of functions. We introduce the product (star-product)
of two functions $f_{\hat A}({\bf x})$ and $f_{\hat B}({\bf x})$
corresponding to two operators $\hat A$ and $\hat B$, respectively, by
the relations
\begin{equation}\label{eq.5}
f_{\hat A\hat B}({\bf x})=f_{\hat A}({\bf x})* f_{\hat B} ({\bf
x})=\mbox{Tr}\,(\hat A\hat B\hat{\cal U}({\bf x})).
\end{equation}
The standard product of the operators in the Hilbert space is
the associative product,
$\hat A(\hat B \hat C)=(\hat A\hat B)\hat C$,
then the star-product of functions -- symbols of operators has to be
associative too
\begin{equation}\label{eq.6}
f_{\hat A}({\bf x})*(f_{\hat B}({\bf x})*f_{\hat C}({\bf x}))=
(f_{\hat A}({\bf x})*f_{\hat B}({\bf x}))*f_{\hat C}({\bf x}).
\end{equation}
The map provides the nonlocal product of two functions
(star-product)
\begin{eqnarray*}
f_{\hat A}({\bf x})*f_{\hat B}({\bf x})=\int f_{\hat A}({\bf
x}'')f_{\hat B}({\bf x}') K({\bf x}'',{\bf x}',{\bf x})\,d{\bf
x}'\,d{\bf x}''.
\end{eqnarray*}
The kernel of star product is linear with respect to the
dequantizer and nonlinear in the quantizer operator
\begin{eqnarray*}
K({\bf x}'',{\bf x}',{\bf x})= \mbox{Tr}\left[\hat{\cal D}({\bf
x}'')\hat{\cal D}({\bf x}') \hat{\cal U}({\bf x})\right].
\end{eqnarray*}
The associativity condition for operator symbols means that the
kernel of star-product of symbols of operators $K({\bf x}'',{\bf
x}',{\bf x})$ satisfies the nonlinear equation \cite{Patricia}
\begin{equation}\label{patC2}
\fl \int K({\bf x}_1,{\bf x}_2, {\bf y})K({\bf y},{\bf x}_3, {\bf
x}_4)d{\bf y}=\int K({\bf x}_1,{\bf y}, {\bf x}_4)K({\bf x}_2,{\bf
x}_3, {\bf y})d{\bf y}.
\end{equation}

Now consider, following \cite{Patricia}, another scheme
\begin{equation}\label{eqp1}
f^{(d)}_{\hat A}({\bf x})=\mbox{Tr}\left[\hat A\hat{\cal D}({\bf
x})\right], \quad  \hat A= \int f^{(d)}_{\hat A}({\bf x})\hat{\cal
U}({\bf x})~d{\bf x}.
\end{equation}
We replace the quantizer and dequantizer by each other
because the compatibility condition is valid in the both cases.
We consider the quantizer--dequantizer pair as dual to the initial one
\begin{eqnarray*}
\hat{\cal U}^\prime({\bf x})= \hat{\cal D}({\bf x}),\quad \hat {\cal
D}^\prime({\bf x})=\hat{\cal U}({\bf x}),
\end{eqnarray*}
The interchange corresponds to a specific symmetry of the equation
for associative star-product kernel. The star-product of dual
symbols~$f^{(d)}_{\hat A}({\bf x})$, $f^{(d)}_{\hat B}({\bf x})$ of
two operators $\hat A$ è $\hat B$ is described by dual integral
kernel
\begin{eqnarray*}
K^{(d)}({\bf x}'',{\bf x}',{\bf x})=\mbox{Tr}\left[\hat{\cal U}({\bf
x}')\hat{\cal U}({\bf x}'')\hat{\cal D}({\bf x})\right],
\end{eqnarray*}
The dual kernel is another solution of nonlinear equation
(\ref{patC2}).

\section{Symplectic tomography}
We consider symplectic tomography scheme \cite{Mancini1} as an
example of the star-product quantization, following
\cite{Marmo,Kapusta}. In the symplectic-tomography scheme, the
tomographic symbol $f_{\hat A}({\bf x})$ of the operator $\hat A$ is
obtained by means of the dequantizer
\begin{eqnarray*}
\hat{\cal U}(X,\mu,\nu)=\delta(X\hat 1-\mu\hat q-\nu\hat p)
\end{eqnarray*}
where vector ${\bf x}=(X,\mu,\nu)$ has the coordinates which are
real numbers and $\hat 1$ is identity operator. The quantizer in
symplectic tomography reads
\begin{eqnarray*}
\hat{\cal D}(X,\mu,\nu)=\frac{1}{2\pi} \exp\left(iX\hat1-i\nu\hat
p-i\mu\hat q\right).
\end{eqnarray*}
The kernel of star-product of two tomographic symbols of operators
$\hat A$ and $\hat B$  has the form \cite{Marmo}
\begin{eqnarray*}
&&K(X_1,\mu_1,\nu_1,X_2,\mu_2,\nu_2,X\mu,\nu)=
\frac{\delta\Big(\mu(\nu_1+\nu_2)-\nu(\mu_1+\mu_2)\Big)}{4\pi^2}\nonumber\\
&&\times\exp\Big(\frac{i}{2}\Big\{\left(\nu_1\mu_2-\nu_2\mu_1\right)\Big.\Big.
\Big.\Big.+2X_1+2X_2-\frac{2(\nu_1+\nu_2)X}{\nu}\Big\}\Big).
\end{eqnarray*}
It is worth noting the important property of the described tomographic map.
If the operator under consideration is a density
operator $\hat\rho$, its tomographic symbol $w(X,\mu,\nu)$ is the
standard probability density of continuous real variable $X$, i.e.,
the function called symplectic tomogram of the state
$w(X,\mu,\nu)=\mbox{Tr}\,\hat\rho\delta(X\hat 1-\mu\hat q-\nu\hat p).$
It is nonnegative and normalized
$\int w(X,\mu,\nu)dX=1.$
The mean value of the quantum observable $\hat A$ reads
\begin{eqnarray*}
&&\langle \hat A\rangle=\mbox{Tr}\big(\hat\rho\hat A\big)=
\mbox{Tr}\int w\big(X,\mu,\nu\big)\hat{\cal D}(X,\mu,\nu)\hat A\, d
X\,d\mu\, d\nu.
\end{eqnarray*}
Having in mind that the dual symbol is determined by formulae
(\ref{eqp1}), we obtain
\begin{eqnarray*}
&&\langle\hat A\rangle= \int w\big(X,\mu,\nu\big)f_{\hat
A}^{(d)}(X,\mu,\nu)\, d X\,d\mu\, d\nu.
\end{eqnarray*}
Thus, the mean value of an observable $\hat A$ is given by the
integral of the product of the tomographic symbol of the density
operator and the symbol of the observable in the dual scheme. In
fact, it can be shown that this observation is true in general.

\section{Classical tomographic symbols}
The reversible relationship between the tomographic symbol
$w_f(X,\mu,\nu)$ of the probability distribution $f(q,p)$ in
classical mechanics is determined by formulae (\ref{eq.1a}) and
(\ref{eq.1c}) (see, for example, \cite{MankoManko}). Here we assumed
for $f(q,p)$ the normalization condition $ \displaystyle{\frac{1}{2\pi}}\int
f(q,p)\,dq\,dp = 1$
analogously to the normalization condition for the Wigner function
$W(q,p)$. According to~\cite{MankoManko}, in classical mechanics one
can introduce the operators $\hat A_{\rm cl}$ for which their formal
Weyl symbol $W_{A_{\rm cl}}(q,p)$ coincides with a classical
observable $A(q,p)$, i.e.,
\begin{equation}\label{wf}
W_{A_{\rm cl}}(q,p)=2\,\mbox{Tr}\,\hat A_{\rm cl}\hat D(2\alpha)\hat
I=A(q,p).
\end{equation}
The classical tomographic symbol for the observable $A(q,p)$ is the
same as its quantum tomographic symbol, if the Weyl symbol
$W_{A}(q,p)$ for the observable $\hat A$ and the phase-space
function $A(q,p)$ coincide. Really, expressions~(\ref{wf}) also
represent the relationship between tomoghraphic symbol of the
observable $\hat A$ in quantum mechanics and its Weyl symbol
$W_{A}(q,p)$ \cite{MankoManko}. Consequently, we can consider the
phase-space function $A(q,p)$ as the classical Weyl symbol of the
observable $A(q,p)$ in classical mechanics. For example, the quantum
tomographic symbol for a unity operator \cite{Shchukin} and
classical tomographic symbol for unity operator \cite{Pilyavets} are
\begin{eqnarray*}
w_1(X,\mu,\nu)=-\pi|X|\delta(\mu)\delta(\nu). \label{w1}
\end{eqnarray*}
\noindent Since in quantum mechanics Weyl symbols for the position
operator $\hat q$ and momentum operator $\hat p$ are $c$-numbers $q$
and $p$, we have for both classical and quantum tomographic symbols
\begin{eqnarray*}
w_q(X,\mu,\nu)=
\frac{\pi}2X|X|\delta'(\mu)\delta(\nu),
 \quad w_p(X,\mu,\nu)
=\frac{\pi}2X|X|\delta(\mu)\delta'(\nu). \nonumber
\end{eqnarray*}
The commutative star-product kernel
$K(X,\mu,\nu,X_1,\mu_1,\nu_1,X_2,\mu_2,\nu_2)$ for two classical
tomographic symbols $w_{f1}(X_1,\mu_1,\nu_1)$ and
$w_{f2}(X_2,\mu_2,\nu_2)$ was defined in~\cite{MankoManko}. It is of
the form
\begin{eqnarray*}
\fl K(X,\mu,\nu,X_1,\mu_1,\nu_1,X_2,\mu_2,\nu_2) &=&
\frac{1}{(2\pi)^2}e^{i\left(X_1+X_2-X({\nu_1+\nu_2})/{\nu}\right)}
\delta\Big(\nu(\mu_1+\mu_2)-\mu(\nu_1+\nu_2)\Big).
\end{eqnarray*}
The relationship between tomographic star-product kernels in quantum
and classical mechanics reads \cite{Pilyavets}
\begin{eqnarray*}
\fl K_{\mbox{\scriptsize{quant}}}(X,\mu,\nu,X_1,\mu_1,\nu_1,X_2,\mu_2,\nu_2)
=
K_{\mbox{\scriptsize{cl}}}(X,\mu,\nu,X_1,\mu_1,\nu_1,X_2,\mu_2,\nu_2)
e^{[i\left(\mu_2\nu_1-\mu_1\nu_2\right)/2]}.
\end{eqnarray*}

\section{Conclusions}
To conclude, we formulate the main results of our study.
We reviewed the generic approach of
constructing operator symbols and their star-product.
We have shown that the probability representation of quantum states can be
constructed using specific versions of star-product schemes.
In the star-product scheme, the physical interpretation of dual structures is
shown on the example of symplectic tomography.

\section*{Acknowledgements}
The author thank the Organizers of the XV Central European Workshop
on Quantum Optics (Belgrade, May-June, 2008) for kind hospitality
and the Russian Foundation for Basic Research for Travel Grant
No.~08-02-08125.

\end{document}